\documentclass{ws-procs9x6}

\begin{document}


\title{Shell structures in oxygen isotopes described with
modern nucleon-nucleon interactions}

\author{S. Fujii}
\address{Department of Physics, University of Tokyo\\
Tokyo 113-0033, Japan\\
E-mail: sfujii@nt.phys.s.u-tokyo.ac.jp}

\author{R. Okamoto and K. Suzuki}
\address{Department of Physics, Kyushu Institute of Technology\\
Kitakyushu 804-8550, Japan}  


\maketitle

\abstracts{
Shell structures in the $N\simeq Z$ nucleus $^{17}$O and the neutron-rich
oxygen isotopes $^{23}$O and $^{25}$O are microscopically described
by calculating single-particle energies
with modern nucleon-nucleon interactions within the framework of
the unitary-model-operator approach.
It is found that the effect of three-body cluster terms on the single-particle
energy is more important in $^{23}$O and $^{25}$O than $^{17}$O.
}


\section{Introduction}

The shell structure which is closely related
to the single-particle level is one of the fundamental properties in nuclei.
Recently, it has been argued that the shell structure
in neutron or proton-rich nuclei is different from that in stable nuclei,
and thus, some magic numbers disappear
and new magic numbers arise in nuclei
near the drip lines.\cite{Ozawa00,Otsuka01}
When we calculate the energies of single-particle levels in neutron- or
proton-rich nuclei, it would be desirable that the calculation formalism
is based on the particle basis.
Advantages of the particle-basis formalism are that the
Coulomb force can be treated accurately for the $pp$ channel and
effects of charge dependence in realistic nuclear forces
are taken into account in structure calculations.
In the particle-basis formalism, one can obtain the energy
differences between proton and neutron levels for not only $N\simeq Z$
nuclei but also neutron- or proton-rich nuclei in the same way.

As one of the methods for solving nuclear many-body problems,
we have developed a many-body theory,
the unitary-model-operator approach (UMOA).\cite{Suzuki94}
In the UMOA, an energy-independent and Hermitian effective interaction is
derived through a unitary transformation of an original Hamiltonian.
The unitary transformation for the construction of the Hermitian effective
interaction has also been performed
in the no-core shell-model calculation.\cite{Navratil01}
Recently, we have extended the formulation of the UMOA from the isospin basis
to the particle one, and actually applied to the calculations of
binding energies and single-particle energies of the neutron and proton
in nuclei around the $N=Z$ nucleus $^{16}$O
with modern nucleon-nucleon interactions.\cite{Fujii03}
In that work, the effect of two-particle one-hole ($2p$-$1h$) excitation from
the ground state of $^{16}$O has been taken into account in the calculation of
single-particle energies in $^{17}$O and $^{17}$F, but higher-order many-body
correction terms such as three-body cluster (3BC) terms
have not been evaluated.

In the present study, we apply the extended UMOA to neutron-rich oxygen isotopes
in addition to $^{17}$O, and calculate single-particle energies
with the 3BC correction terms.
As far as we know, the present study is the first attempt to clarify
the 3BC effect on the single-particle energy in neutron-rich nuclei
with a realistic nucleon-nucleon interaction.
In the present work, we use the CD-Bonn potential\cite{Machleidt96}
as the realistic nucleon-nucleon interaction.
In the following, after the calculation procedure is briefly presented,
the calculated results are shown.

\section{Calculation procedure}

Since the methods for determining the effective interaction and calculating
the single-particle energy have been given in detail
in the previous work,\cite{Fujii03}
we here outline the calculation procedure.
The harmonic-oscillator (h.o.) wave functions are used
as the basis states in the calculations.
We consider a model space for each of the $nn$, $np$, and $pp$ channels
specified by a boundary number $\rho _{1}$
which is given with the sets of h.o. quantum numbers $\{ n_{a},l_{a}\}$
and $\{ n_{b},l_{b}\}$ of two-body states by
\begin{equation}
\label{rho1}
\rho _{1}=2n_{a}+l_{a}+2n_{b}+l_{b}.
\end{equation}
The value of $\rho _{1}$ is taken as large as possible so that
the calculated results do not depend on this value.
In the present study, we take as $\rho _{1}=12$ as a sufficiently large value.
As for the value of the h.o energy $\hbar \omega$ we use the optimal value
for each nucleus, which has been discussed in the previous work.\cite{Fujii03}

In the large model space, we calculate the two-body effective interaction
self-consistently with the single-particle potential.
The two-body effective interaction in the large model space
is determined so that there are no vertices which induce
$2p$-$2h$ excitation.
Using the two-body effective interaction,
we calculate the unperturbed ground-state energy.
Furthermore, the $2p$-$1h$ ($1p$-$1h$) effect on a nucleus
having a single-particle structure (a closed-shell nucleus) is obtained
by the diagonalization of the unitarily transformed Hamiltonian
which contains the kinetic and single-particle potential energies
and the two-body effective interaction.
The ground-state energy is given by the sum of
the unperturbed ground-state energy and the correlation energy.
In the present study, we further evaluate the 3BC effects both
on the single-particle system and the closed-shell nucleus,
and then the energy of the 3BC effect is added to the ground-state energy.
The full details of the 3BC terms for the closed-shell nucleus
and the single-particle state based on the particle basis
will be shown in a separated paper.
The single-particle energy for the ground state
is expressed as the difference of the ground-state energies of
a nucleus having the single-particle structure
and the corresponding closed-shell nucleus.
Note that the values of $\hbar \omega$ for these two nuclei are,
in general, different from each other
since we search for the optimal $\hbar \omega$ for each nucleus.

\section{Results and discussion}

In the left side of Fig.~1, the calculated results of single-particle energies
for the ground states of $^{17}$O, $^{23}$O, and $^{25}$O
using the CD-Bonn potential are illustrated.
The results are shown separately for the cases with and without the 3BC effect.
It is seen that the 3BC effect has significant contributions repulsively
to the single-particle energies and becomes larger as the mass number becomes
larger.
The strong repulsive effect is favorable for the neutron-rich nuclei
$^{23}$O and $^{25}$O to reproduce the experimental values.
On the contrary, the result for $^{17}$O becomes worse in comparison with the
experimental value when the 3BC effect is taken into account.
We have also observed that the results
for the N$^{3}$LO potential\cite{Entem03} based on chiral perturbation theory
show the same tendency with those for the CD-Bonn potential.

It should be noted that a genuine three-body
force is not included in the present calculation.
The genuine three-body force would show attractive effects
on the single-particle energies,
and would have different contributions between
the $N\simeq Z$ and neutron-rich nuclei
because of the isospin and density dependences.
We should include the genuine three-body force in the structure calculation
to obtain more reliable results.

\begin{figure}[t]
\begin{center}
\includegraphics[scale=0.82]{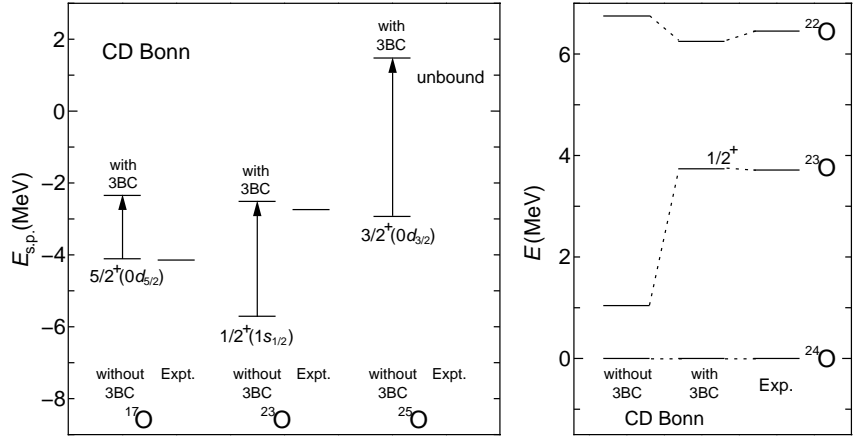}
\end{center}
\caption{Calculated and experimental single-particle energies
for the ground-states of $^{17}$O, $^{23}$O, and $^{25}$O (left figure),
and the energies of the ground states of $^{22}$O, $^{23}$O, and $^{24}$O
relative to $^{24}$O (right figure).
The CD-Bonn potential is employed.}
\label{fig1}
\end{figure}

In order to discuss the importance of the 3BC contribution to
the single-particle energy in neutron-rich oxygen isotopes,
in the right side of Fig.~1,
the energies of the ground-states of $^{22}$O, $^{23}$O, and
$^{24}$O relative to the ground state of $^{24}$O are exhibited.
At the lowest order, $^{22}$O and $^{24}$O are considered to be the
closed-shell nuclei in which the 0$d_{5/2}$ and 1$s_{1/2}$ states of
the neutron are the uppermost occupied states, respectively.
We evaluate the 3BC effect on these closed-shell nuclei.
The ground state of $^{23}$O can be described as the one neutron plus
the ground state of $^{22}$O at the lowest order.
The 3BC terms for the single-particle state of the neutron are calculated.
We see that the 3BC effects on the closed-shell nuclei $^{22}$O and $^{24}$O
do not show significant contributions to the relative energy
between $^{22}$O and $^{24}$O.
The calculated results for $^{22}$O and $^{24}$O are in good agreement with
the experimental values for both the cases with and without the 3BC effect.
On the other hand, as for $^{23}$O, the 3BC effect on the
single-particle state is considerably large, and then the good agreement with
the experimental value is observed when the 3BC effect is taken into account.

\section{Conclusions}

We have applied the extended UMOA based on the particle basis
to the calculation of the single-particle energies of
the ground states in oxygen isotopes with the CD-Bonn potential.
We have evaluated the 3BC effects on closed-shell nuclei
and single-particle states.
The calculated results have shown that the 3BC effect on
the single-particle energy is repulsive and more important
in neutron-rich oxygen isotopes than $^{17}$O.
The strong repulsive effect is favorable for the neutron-rich nuclei
$^{23}$O and $^{25}$O to reproduce the experimental values.
We conclude that the 3BC effect is indispensable for the microscopic
description of the single-particle energies in neutron-rich oxygen isotopes,
as far as the present particle-hole formalism is employed.
The detailed results of the 3BC effect in oxygen isotopes
will be reported elsewhere in the near future.

\section*{Acknowledgments}

One of the authors (S.~F.)
acknowledges the Special Postdoctoral Researchers Program of RIKEN.
This work is supported by a Grant-in-Aid for Scientific Research (C) from
Japan Society for the Promotion of Science (JSPS) (No. 15540280).


\begin{thebibliography}{0}






\bibitem{Ozawa00}
A. Ozawa, T. Kobayashi, T. Suzuki, K. Yoshida, and I. Tanihata,
{\it Phys. Rev. Lett.} {\bf 84}, 5493 (2000).

\bibitem{Otsuka01}
T. Otsuka, R. Fujimoto, Y. Utsuno, B. A. Brown, M. Honma, and T. Mizusaki,
{\it Phys. Rev. Lett.} {\bf 87}, 082502 (2001).

\bibitem{Suzuki94}
K. Suzuki and R. Okamoto, {\it Prog. Theor. Phys.} {\bf 92}, 1045 (1994).

\bibitem{Navratil01}
P. Navr\'atil, J. P. Vary, W. E. Ormand, and B. R. Barrett,
{\it Phys. Rev. Lett.} {\bf 87}, 172502 (2001).

\bibitem{Fujii03}
S. Fujii, R. Okamoto, and K. Suzuki, {\it nucl-th}/0311053.

\bibitem{Machleidt96}
R. Machleidt, F. Sammarruca, and Y. Song,
{\it Phys. Rev.} {\bf C53}, R1483 (1996).

\bibitem{Entem03}
D. R. Entem and R. Machleidt, {\it Phys. Rev.} {\bf C68}, 041001(R) (2003).



\end{thebibliography}
\end{document}